\documentclass[%
phys,
jmp,
amsmath,amssymb,
reprint,
author-numerical,%
sort,
]{revtex4-1}

\usepackage{graphicx}
\usepackage{dcolumn}
\usepackage{bm}
\usepackage{multirow}

\begin{document}


\title{Modular nanomagnet design for spin qubits confined in a linear chain}

\author{Michele Aldeghi}
\email{mia@zurich.ibm.com}
\author{Rolf Allenspach}%
\author{Gian Salis}%
\affiliation{%
 IBM Research-Zurich, Säumerstrasse 4, 8803 Rüschlikon, Switzerland}%

\date{\today}

\begin{abstract}

On-chip micromagnets enable electrically controlled quantum gates on electron spin qubits. Extending the concept to a large number of qubits is challenging in terms of providing large enough driving gradients and individual addressability. Here we present a design aimed at driving spin qubits arranged in a linear chain and strongly confined in directions lateral to the chain. Nanomagnets are placed laterally to one side of the qubit chain, one nanomagnet per two qubits. The individual magnets are "U"-shaped, such that the magnetic shape anisotropy orients the magnetization alternately towards and against the qubit chain even if an external magnetic field is applied along the qubit chain. The longitudinal and transversal stray field components serve as addressability and driving fields. Using micromagnetic simulations we calculate driving and dephasing rates and the corresponding qubit quality factor. The concept is validated with spin-polarized scanning electron microscopy of Fe nanomagnets fabricated on silicon substrates, finding excellent agreement with micromagnetic simulations. Several features required for a scalable spin qubit design are met in our approach: strong driving and weak dephasing gradients, reduced crosstalk and operation at low external magnetic field. 
\end{abstract}

\keywords{Nanomagnets, electron spin qubit, EDSR, finFET, nanowire}
\maketitle

Single-qubit gates in electron-spin qubits are typically implemented with electron dipole spin resonance (EDSR) enabled by a magnetic field gradient~\cite{watson, Kawakami2014, takeda2016fault, Pioro-Ladriere2008}. Successful operation of up to 6 qubits with a common micromagnet has been demonstrated~\cite{philips2022universal}. 
Despite its current success, scaling this technique to drive and address thousands of single qubits remains challenging~\cite{ gonzalez2021scaling, tadokoro2021designs, boter2021spider, nakamura2022micromagnet,  yoneda2015robust} because the magnetic stray field needs to fulfill different requirements at the same time. It should provide: (i) large driving gradients to achieve a fast Rabi frequency $f_{\mathrm{Rabi}}$, (ii) a difference in Larmor frequencies $f_{\mathrm L}$ between neighboring qubits significantly larger than $f_{\mathrm{Rabi}}$ for qubit addressability with low crosstalk and (iii) small gradients of its longitudinal field component to prevent spin dephasing.

A possible strategy to address this challenge is to place individual nanomagnets close to the quantum dots that define the qubits, e.g. by co-integration of the magnets with the hosting structure. For this purpose, platforms offering an inherent confinement into a one-dimensional geometry such as Si nanowires~\cite{froning2018single} or finFETs~\cite{kuhlmann2018ambipolar} are of advantage, since less conventional gate electrodes are needed to control the qubits, freeing up space to place the nanomagnets. In the finFET platform, electron spin qubits have been realized by a CMOS fab-compatible process~\cite{zwerver2022qubits}. Due to the strong lateral confinement induced by the fin, the displacement of the electron wavefunction for EDSR driving is preferably induced along the fin, i.e. along the qubit chain. This is different compared to currently established micromagnet designs, where a displacement transverse to the qubit chain is used to drive EDSR~\cite{yoneda2015robust, philips2022universal, zwerver2022qubits}. 

\begin{figure}[ht]
	\centering
	\includegraphics[width=0.48\textwidth]{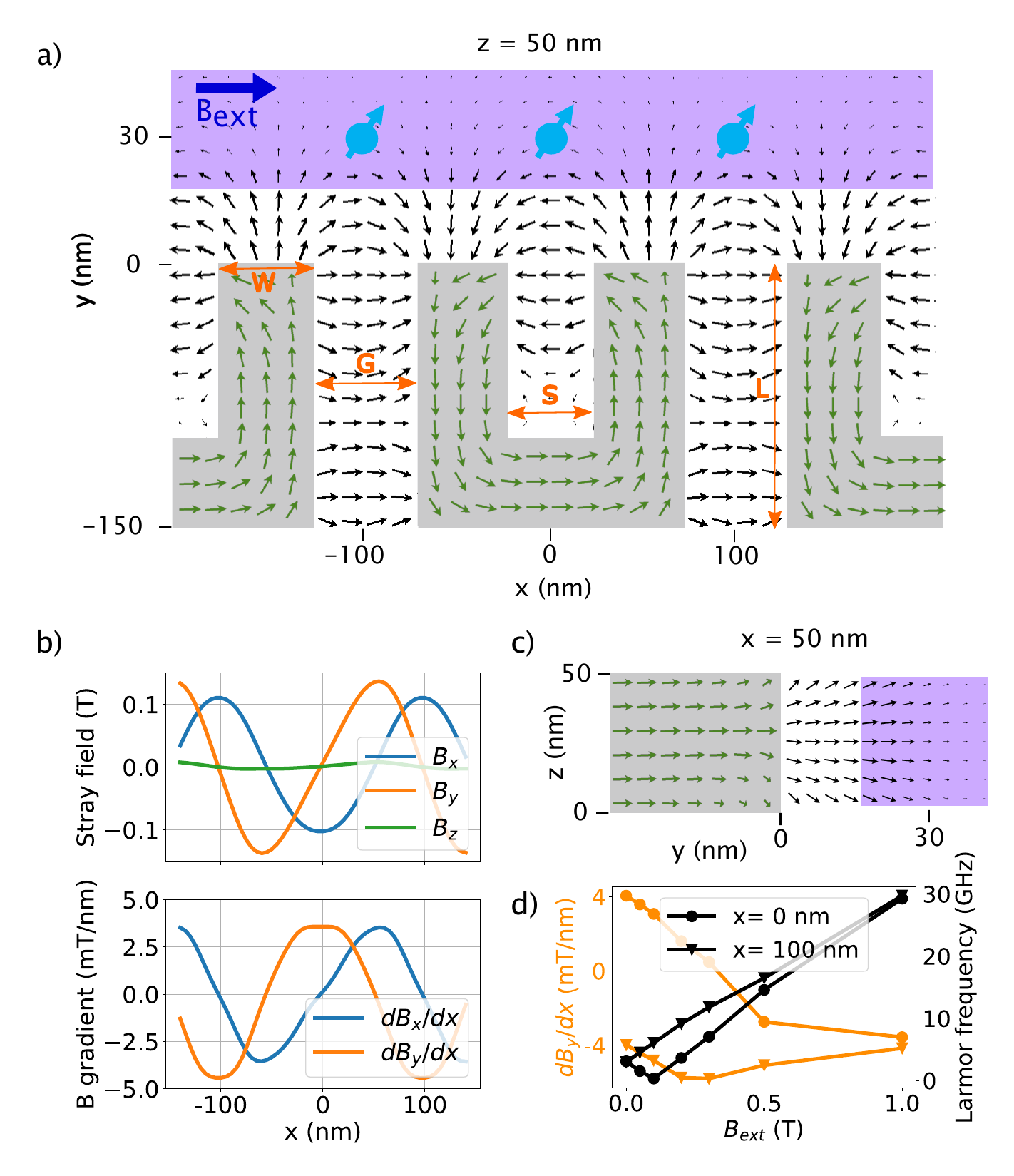}
	\caption{\label{fig:fig_1} (a) Nanomagnet design for a linear chain of qubits with 1D confinement (violet). An external field $B_{\mathrm{ext}}= 0.05$\,T is applied along $x$ such that the arms of the nanomagnets (grey) are magnetized alternatively into $+y$ and $-y$ directions.  EDSR driving is performed by displacing the electron wavefunction of a quantum dot (light blue) along the $x$ direction. The calculated stray field $B$ of the nanomagnets (black arrows) and magnetization pattern (green arrows) are overlaid. (b) Line cut showing stray field and stray field gradient along the qubit chain at $z=25$\,nm. The maxima of the driving gradient ($dB_y/dx$) coincide with the zero crossings of the dephasing gradient ($dB_x/dx$), setting the so-called sweet spot. (c) Cross-section of calculated magnetization pattern at $x= 50$\,nm. (d) Larmor frequency and driving gradient vs. $B_{\mathrm{ext}}$ at the two different qubit locations at $z=25$\,nm.}
\end{figure}

In this letter we present a modular design of nanomagnets aimed at driving a large number of spin qubits arranged in a linear chain and strongly confined in directions lateral to the chain. We have simulated and fabricated "U"-shaped Fe nanomagnets down to a feature size of $50$\,nm. Their magnetization pattern has been imaged by spin-polarized scanning electron microscopy (spin-SEM)~\cite{allenspach2000spin}, showing excellent agreement with micromagnetic simulations. We discuss sweet spots for reduced influence of charge noise and reduced crosstalk in EDSR driving and investigate how these properties depend on the alignment accuracy between nanomagnets and the spin qubits. The concept presented provides a scalable approach to EDSR driving of electron spins and can be implemented in several spin qubit platforms ranging from FinFETs to nanowires and planar geometries.

Figs.~\ref{fig:fig_1}(a) and (c) show top and side views of our design. Individual "U"-shaped nanomagnets are placed laterally to one side of the qubit chain, one nanomagnet per two qubits. The two ends of the "U" point towards the qubit chain. Because of magnetic shape anisotropy, the magnetization in the two arms of each nanomagnet points in an alternating way towards and against the chain even with an external magnetic field $B_{\mathrm{ext}}$ applied along the qubit chain. This creates a gradient in the stray field component transverse to the external field. This gradient is used to drive Rabi oscillations of the spin qubits by spatially oscillating the position of the host electron in the direction along the chain. The sign of the longitudinal component of the stray field alternates from qubit to qubit, which together with $B_{\mathrm{ext}}$ provides large differences in the Larmor frequencies of neighboring qubits and enables addressability with low crosstalk.  

Micromagnetic simulations have been performed with the software package MuMax$^3$~\cite{vansteenkiste2014design}. We choose Fe as the magnetic material and assume a saturation magnetization of 1.7 MA/m, an exchange stiffness of 21 pJ/m, a cell size of 1 nm and zero temperature. The $x-y$ plane defines the sample surface, with the qubit chain oriented along the $x$ direction. The nanomagnets have an arm width $W= 50$\,nm, length $L= 150$\,nm, thickness $T= 50$\,nm, separation $S= 50$\,nm between the arms and $G= 50$\,nm between the magnets, which results in a qubit-qubit pitch of $100$\,nm. Smaller pitches can be implemented by decreasing the size of $S$ and $G$, which would further increase the attainable driving gradient at the qubit locations.

The stray field $\mathbf{B}=(B_x, B_y, B_z)$ of the nanomagnet at $B_{\mathrm{ext}}=0.05$\,T is represented in a quiver plot in Fig.~\ref{fig:fig_1}(a) and (d). Its longitudinal ($x$ component) maxima and minima coincide with the zero-crossings of its transverse ($y$) component [Fig.~\ref{fig:fig_1}(b)], setting optimal locations for the qubits at the center of the nanomagnets $x= 0$\,nm ("arm") and in between them at $x= 100$\,nm ("gap"). At $z= 25$\,nm and $y= 30$\,nm, the driving gradient $dB_{y}/dx$ reaches 3.9 mT/nm at the "arm" position and -4.8 mT/nm at the "center" position, see also Table~\ref{table:tab_1}. Together with a zero-crossing of the dephasing gradient $dB_{x}/dx$, these positions define a sweet spot.

The electric field that induces the oscillatory spatial displacement of the electron wavefunction of an individual quantum dot will unavoidably also affect neighboring qubits. Individual addressability therefore requires qubit frequencies $f_{\mathrm L}$ to be separated far enough to only resonantly excite one qubit, which translates into having distinct magnetic fields among neighboring qubits. At neighboring qubit positions, $B_x$ has opposite signs, and together with a finite $B_{\mathrm{ext}}$, $f_{\mathrm L}$ of these qubits will be different. We find that at $B_{\mathrm{ext}}= 0.05$\,T, $f_{\mathrm L}= 1.84$\,GHz ("arm") and $4.64$\,GHz ("gap"), assuming an electron g-factor of $2.0$ [Fig.~\ref{fig:fig_1}(d)]. When set to 0.10\,T, $B_{\mathrm{ext}}$ matches $|B_x|$ such that $f_{\mathrm L} \approx 0$ at the "arm" location. For $B_{\mathrm{ext}}>0.3$\,T, the driving gradient decreases significantly due to the reorientation of the magnetization pattern in the arms. At $B_{ext}=2.0$\,T, the magnetization almost saturates everywhere into the $x$ direction and neighbouring qubits approach the same Larmor frequency. The magnitude of the driving gradient then approaches the values at low fields.  

Hence, in our design $B_{\mathrm{ext}}$ has to be adjusted such that its magnitude is as low as possible to preserve a high driving gradient and at the same time high enough to allow for single qubit addressability. These values for $B_{\mathrm{ext}}$ can be significantly smaller than what is commonly used in EDSR experiments and simulations with micromagnets~\cite{watson, Kawakami2014, takeda2016fault, Pioro-Ladriere2008, philips2022universal,  struck2020low, yoneda2018quantum, neumann2015simulation, chesi2014single}. 

Several advantages arise at low values of $B_{\mathrm{ext}}$. Firstly, a lower $f_{\mathrm L}$ leads to reduced crosstalk between neighboring drive lines, due to the frequency dependence of their capacitive coupling~\cite{tesker2022flopping}. Secondly, less demanding control electronics is required. Thirdly, the spin relaxation time is increased~\cite{glavin2003spin}, also avoiding hot spots when the Zeeman energy coincides with the valley splitting~\cite{huang2014spin}. Finally, coherent shuttling of qubits is facilitated at low fields, since the acquired spin phase during shuttling is directly dependent on the magnitude of $B_{\mathrm{ext}}$~\cite{langrock2022blueprint}.

\begin{figure}[ht]
	\centering
	\includegraphics[width=0.46\textwidth]{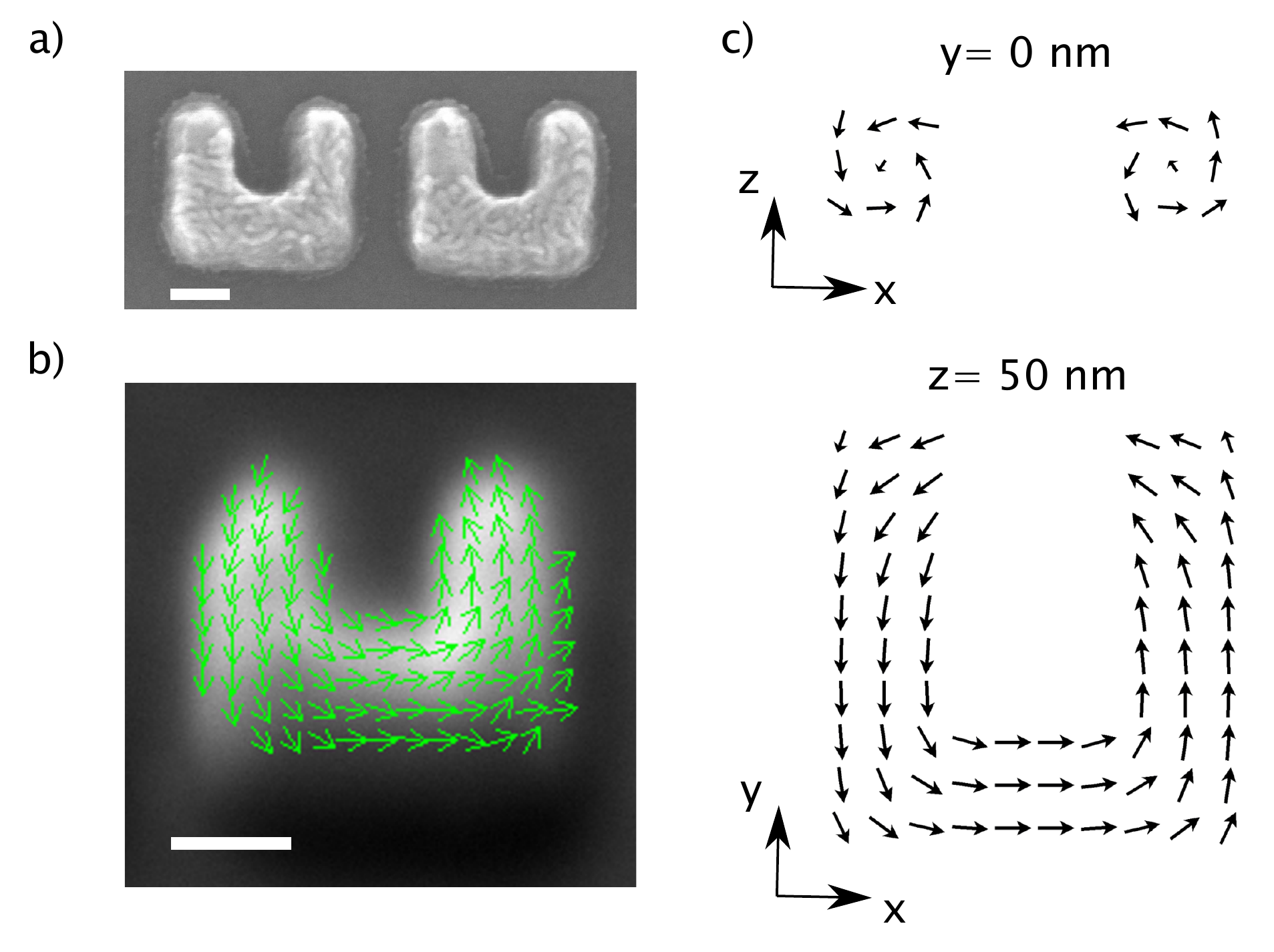}
	\caption{\label{fig:fig_2} (a) SEM image of 50 nm thick Fe nanomagnets. (b) Magnetization pattern measured by spin-SEM, after a magnetic pulse of 17\,mT applied along $+x$. The arrows are obtained from two orthogonal in-plane magnetization components and cropped according to the topographic data. (c) Micromagnetic simulation at $B_{\mathrm{ext}}= 0$\,T. The scale bars in (a) and (b) are 50 nm.}
\end{figure}

The design summarized in Fig.~\ref{fig:fig_1} has four main advantages with respect to previous micromagnet designs: enhanced addressability between neighboring qubits at low external field (difference in $f_{\mathrm L}$ is above 2.5 GHz at $B_{\mathrm{ext}}= 0.05$\,T), stronger driving gradient ($dB_{y}/dx= 3.9$\,mT/nm), a zero-crossing in the dephasing gradient $dB_{x}/dx$ and a modular design that can be extended to arbitrarily long chains of spin qubits.

We have fabricated such "U"-shaped nanomagnets on a silicon substrate covered by native oxide, using e-beam lithography on a PMMA-based bilayer followed by e-beam evaporation and lift-off of 50\,nm of Fe and 3\,nm of Pd as a capping layer. Before the magnetic measurements, the capping layer was removed in ultrahigh vacuum by Ne ion sputtering. An in-plane magnetic field pulse of 17\,mT was applied along $+x$ to orient the bases of the magnets along the same direction. The magnets were then investigated in remanence (i.e. at $B_{\textrm{ext}}=0$) by spin-SEM, where topography and magnetization distribution of the first 2 nm of the sample surface are simultaneously determined with a lateral resolution of approximately 10 nm.

In Fig.~\ref{fig:fig_2}(a) we show an SEM image of the fabricated nanomagnets. The patterning process caused the edges to be rounded with respect to the geometry used in the simulations. This rounding is also visible in the spin-SEM image in Fig.~\ref{fig:fig_2}(b), where an arrow-plot combining both in-plane magnetization components is overlaid on top of the topographic data. The magnetization follows the shape of the magnet similar to that obtained in the simulations at $B_{\mathrm{ext}}= 0$\,T shown in Fig.~\ref{fig:fig_2}(c). Deviations between simulation and experimental magnetization pattern are found in the rounded region at the apices of the arms. In Fig.~\ref{fig:fig_2}(b) we see that the magnetization rotates its direction at the apex of both arms along the $x$ direction. This is compatible with the vortex visible in the simulation $x-z$ cut at $y= 0$\,nm in Fig.~\ref{fig:fig_2}(c). 

We investigate the impact of the vortex formation by comparing the simulation described in Fig.~\ref{fig:fig_1} ("relaxed") with a hypothetical nanomagnet with arms uniformly magnetized in opposite directions ("uniform"). We see that the rearrangement of the magnetization pattern reduces the driving gradient as well as $f_{\mathrm L}$ at both qubit positions (Table~\ref{table:tab_1}). 

A smaller stray field is also found in the simulation where nanomagnets have a rounded apex ("rounded'). We approximate the shape of the nanomagnet apex [Fig.~\ref{fig:fig_2}(a)] with a semi-cylinder with a diameter of $50$\,nm. All other simulation parameters are identical to the "relaxed" case. The comparison between these two geometries shows a reduction of the driving gradients by $\approx 20\%$ at both qubit locations for the "rounded" case. This reduction is ascribed to the smaller amount of magnetic material present in the vicinity of the qubit.


\begin{table}[bt]
\centering
\begin{tabular}{ |p{0.9cm}|p{2.7cm}|p{1.45cm}|p{1.35cm}|p{1.45cm}|  }
 \hline
 \multicolumn{2}{|c|}{Position} & uniform & relaxed & rounded\\
 \hline
 \multirow{0}*{gap}  & $dBy/dx$ (mT/nm)& $-5.95$ &  $-4.79$ &$-3.74$\\
 & $f_{\mathrm L}$ (GHz) &$5.98$ &$4.64$ &$4.09$\\
 \multirow{0}*{arm}  & $dBy/dx$ (mT/nm)& $6.05$ &  $3.90$ &$3.14$\\
 & $f_{\mathrm L}$ (GHz) &$3.57$ &$1.84$ &$1.41$\\
 \hline
\end{tabular}
\caption{Driving gradient $dBy/dx$ and Larmor frequency $f_{\mathrm L}$ at the "gap" and "arm" optimal qubit locations, at $y= 30$\,nm and $z= 25$\,nm and $B_{\mathrm{ext}}= 0.05$\,T. In the simulation "uniform", the nanomagnet has a hypothetical uniform magnetization of opposite direction in the two arms. The simulation "relaxed" refers to the simulation described in the main text. The "rounded" simulations differs from "relaxed" by the shape of the nanomagnets, which have been set as a semi-cylinder with diameter $50$\,nm.}
\label{table:tab_1}
\end{table}

\begin{figure}[ht]
	\centering
	\includegraphics[width=0.42\textwidth]{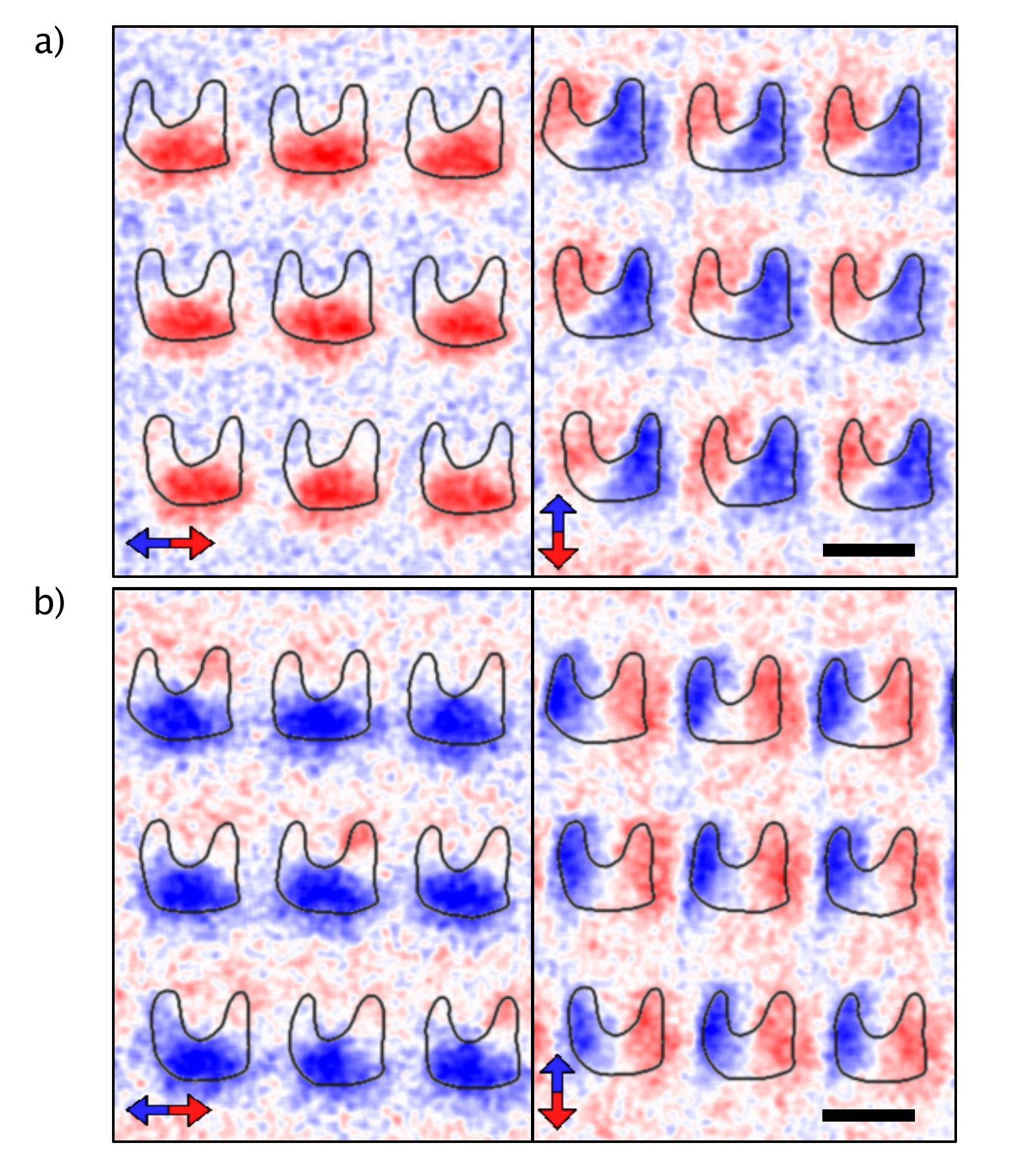}
	\caption{\label{fig:fig_3} Magnetization pattern of an array of nanomagnets after applying a field of 17\,mT along $+x$ (a) and $-x$ (b). The left column of each group shows the magnetization component along the $x$ direction, where the $y$ component is shown in the right one. The scale bar is 150 nm.}
\end{figure}

Scalabe designs of one-dimensional arrays of qubits can be implemented by repeating the nanomagnets along the chain, provided that magnetization patterns among the different nanomagnets are similar. In Fig.~\ref{fig:fig_3} we show both in-plane magnetization components of a 3 by 3 nanomagnet array after applying a magnetic pulse of 17\, mT along $+x$ and $-x$. We extract the nanomagnets contour from the topographic data and overlay it on the magnetic measurement images. The comparison of the magnetization components between the two measurements demonstrates a symmetrical reversal of the magnetization pattern for opposite magnetic field directions, with the bases of the magnets oriented along the direction of the magnetic pulse and the arms alternating between $+y$ and $-y$. Upon closer inspection, we see in the horizontal magnetization images that the magnetization at the apices is partially oriented along the $x$ direction, as already noted in Fig.~\ref{fig:fig_2}(b). The sign of this component varies among the different apices, suggesting a varying chirality of the vortices, both within and between structures. We believe that the chirality is induced by random local defects, and according to the micromagnetic simulations, the energy difference between nanomagnets with identical or opposite vortex chirality is negligible. The vortex chirality has no significant impact on the driving gradient or addressability at the foreseen qubit positions. 

Large driving gradients are accompanied by large dephasing gradients and may limit the fidelity of qubit operations. We therefore calculate the dephasing time and quality factor that we expect from the "U"-shaped nanomagnets. Charge noise has been identified as the main limitation to the gate fidelity of single-qubit gates for micromagnet-based EDSR~\cite{struck2020low, yoneda2018quantum}, and will be considered as the dephasing mechanism in the following. We assume that the charge noise has a 1/f power spectrum~\cite{yoneda2018quantum, nakajima2020coherence, dumoulin2021low}. The noise in electric field will translate into noise in the spatial position of the electron wavefunction and through the dephasing gradient into noise in $f_{\mathrm L}$. The relevant dephasing gradients are $dB_{x}/dx$, $dB_{x}/dy$ and $dB_{x}/dz$, since the $y$ and $z$ stray field components are negligible at the qubit position [Fig.~\ref{fig:fig_1}(b)]. The gradients are calculated in a 50 x 50 nm$^2$ area by fitting the simulated stray field with a quadratic polynomial at each mesh point and taking the first derivative of the fitted polynomial. Higher orders lead to significantly smaller contributions and are neglected in the following. 

The dephasing rate for each direction $i\in{x, y, z}$ is calculated by assuming root-mean-square displacements $\Delta i$ along the $i$ direction~\cite{nakajima2020coherence} with :
\begin{equation}
\Gamma_i=\pi\sqrt{2}\frac{\gamma_e}{h}\frac{dB_{x}}{di}\Delta i
\end{equation}
where $\gamma_e$ is the electron gyromagnetic ratio (g-factor of 2), $h$ Planck's constant and $dB_{x}/di$ the approximated first derivative of the stray field along the $i$ direction. We assume $\Delta x = 5$\,pm~\cite{Kawakami2014} and $\Delta y,\ \Delta z= 0.5$\, pm. The larger $\Delta x$ is justified by the weaker confinement of the quantum dot along that direction. The dephasing time is then $T_2^*=1/\Gamma=1/(\Gamma_x+\Gamma_y+\Gamma_z)$.

\begin{figure}[ht]
	\centering
	\includegraphics[width=0.46\textwidth]{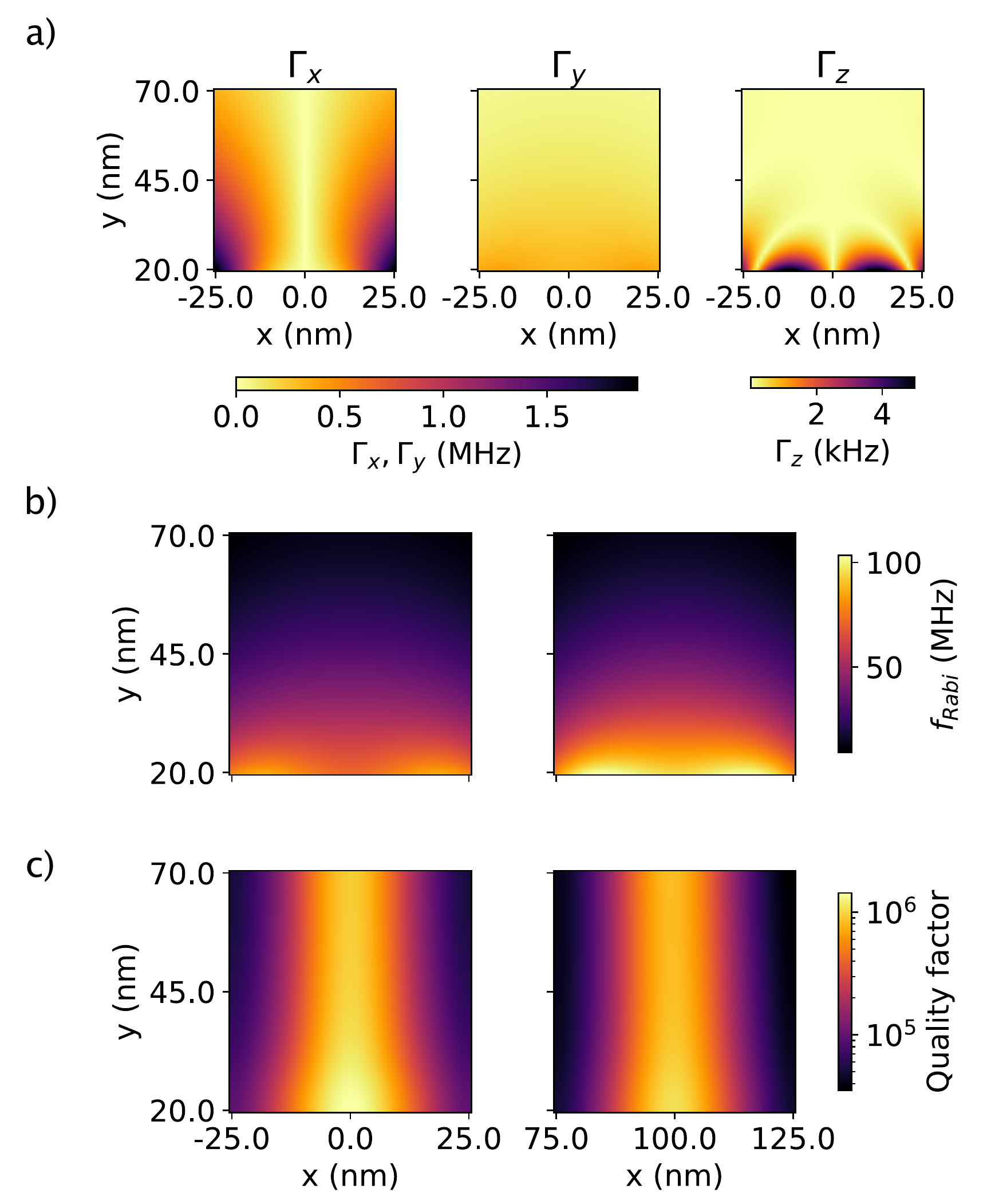}
	\caption{\label{fig:fig_4} Alignment dependencies of (a) dephasing rate, (b) Rabi frequency and (c) quality factor. (a) Shows the dephasing rate induced by charge noise  along the $x$, $y$ and $z$ directions, for qubits placed around the center of the arms of the magnets. The dephasing rates for the qubit at the other location are very similar and not shown here. In (b) and (c) results for qubits placed both at "arm" and "center" positions are shown.}
\end{figure}

Fig.~\ref{fig:fig_4}(a) shows the calculated dephasing rates $\Gamma_x$, $\Gamma_y$ and $\Gamma_z$ as a function of the qubit position around the "arm" position. At $x= O$\,nm, $\Gamma_x$ is zero, and the same is true for the position "gap" at $x= 100$\,nm (data not shown). $\Gamma_y$ provides the main source of dephasing, whereas $\Gamma_z$ is negligible within the simulated range. We note that because the magnetic field is rotation-free outside the nanomagnets $dB_x/dy=dB_y/dx$, and therefore the maxima in the driving field $dB_y/dx$ at the "arm" and "gap" positions coincide with maxima in $\Gamma_y$. Here, the strong confinement of the electron wavefunction in the fin reduces displacements along the $y$ direction, thus mitigating dephasing.

We calculate the quality factor $Q=T_{2}^{\mathrm{Rabi}}/\tau_{\mathrm{Rabi}}$, an upper bound for the qubit fidelity \cite{takeda2016fault}. The Rabi dephasing time is $T_2^{\mathrm{Rabi}}=\pi^2A^2/f_{\mathrm{Rabi}}$~\cite{nakajima2020coherence}, and the Rabi frequency $f_{\mathrm{Rabi}}=1/\tau_{\mathrm{Rabi}}$ is

\begin{equation}
f_{\mathrm{Rabi}}=\frac{\gamma_e}{h}\frac{dB_{y}}{dx}\delta{x}.
\end{equation}

The $A$ factor defines a $1/f$ power noise spectrum $A^2/f$ of the qubit frequency. We obtain $A$ from the relation~\cite{nakajima2020coherence} $\sigma = A\sqrt{log(f_h/f_l)}$, where $f_h$ ($f_l$) is the highest (lowest) frequency of noise the qubit is exposed to, and $\sigma$ is the root-mean-square of the qubit frequency noise. The latter is obtained from $\Gamma$ calculated above by the relation $\Gamma = \pi\sqrt2\sigma$. We assume $f_h= 100$\,MHz and $f_l= 1/3600$\,Hz, similar to the experiment in Ref.~\cite{struck2020low}, and a displacement amplitude $\delta_{x}$ of 500 pm from the EDSR drive, similar to \cite{tesker2022flopping}. Fig. \ref{fig:fig_4}(b) shows the expected Rabi frequency, highlighting an increase in value with decreasing distance between the qubit and the nanomagnet, similar to the stray field trend visible in Fig.~\ref{fig:fig_1}(a).

The quality factor $Q=f_{\mathrm{Rabi}}^2/A^2\pi^2$ is smoothed by a Gaussian filter with a sigma-width of 5\,nm, compatible with a full-width at half maximum of the dot size in $x-y$ plane of 12 nm [Fig.~\ref{fig:fig_4}(c)]. We see that the qubits are optimally placed at $x$ positions aligned with the center of the magnet and the gap between them, corresponding to the dephasing minimum positions described before. The weak dependency of $Q$ on the $y$-coordinate is due to the equality between the driving gradient and the main dephasing gradient ($dB_x/dy$), as a consequence of Ampère's law. On the other hand, we see a strong dependence on the $x$ alignment of the qubit, caused by corresponding increase of $\Gamma_x$. Misalignment of 3 nm in the $x$ direction reduces the quality factor by 30 \%. Nevertheless, quality factors higher than $10^5$ are achieved within 15 nm of misplacement, which exceeds the alignment precision of standard e-beam lithography~\cite{ebl}. Note that the quoted values for $Q$ depend on the assumed displacement noise and scale inversely with $A^2$ and therefore also inversely with the square of the dephasing gradient. 

In conclusion, we have shown a nanomagnet design aimed at driving a linear chain of qubits that are strongly confined in directions lateral to the chain. EDSR can be performed with an external magnetic field below the saturation field of the nanomagnets, which is advantageous for operation of spin qubits at lower frequencies. At the foreseen locations, neighboring qubits differ in Larmor frequency by more than $2.5$\,GHz. We found reproducible magnetization of fabricated nanomagnets and excellent agreement with micromagnetic simulations. The qubits are expected to have have quality factors greater than $10^5$ with alignment tolerances relative to the nanomagnet positions that are within reach for e-beam lithography. The modularity of the design opens up the possibility to address a large number of qubits aligned along a linear chain.

\begin{acknowledgments}
We thank the Cleanroom Operations Team of the Binnig and Rohrer Nanotechnology Center (BRNC), Matthias Mergenthaler, Nico Hendrickx, Felix Schupp for advise on sample processing, Andreas Fuhrer, Patrick Harvey-Collard and Andrea Ruffino for discussions, Eoin Kelly for proof-reading the manuscript and Andreas Bischof and Stephan Paredes for support in the lab. This work was supported as a part of NCCR SPIN funded by the Swiss National Science Foundation (grant number 51NF40-180604).
\end{acknowledgments}

\bibliography{references_nanomagnets.bib}

\begin{thebibliography}{26}%
\makeatletter
\providecommand \@ifxundefined [1]{%
 \@ifx{#1\undefined}
}%
\providecommand \@ifnum [1]{%
 \ifnum #1\expandafter \@firstoftwo
 \else \expandafter \@secondoftwo
 \fi
}%
\providecommand \@ifx [1]{%
 \ifx #1\expandafter \@firstoftwo
 \else \expandafter \@secondoftwo
 \fi
}%
\providecommand \natexlab [1]{#1}%
\providecommand \enquote  [1]{``#1''}%
\providecommand \bibnamefont  [1]{#1}%
\providecommand \bibfnamefont [1]{#1}%
\providecommand \citenamefont [1]{#1}%
\providecommand \href@noop [0]{\@secondoftwo}%
\providecommand \href [0]{\begingroup \@sanitize@url \@href}%
\providecommand \@href[1]{\@@startlink{#1}\@@href}%
\providecommand \@@href[1]{\endgroup#1\@@endlink}%
\providecommand \@sanitize@url [0]{\catcode `\\12\catcode `\$12\catcode
  `\&12\catcode `\#12\catcode `\^12\catcode `\_12\catcode `\%12\relax}%
\providecommand \@@startlink[1]{}%
\providecommand \@@endlink[0]{}%
\providecommand \url  [0]{\begingroup\@sanitize@url \@url }%
\providecommand \@url [1]{\endgroup\@href {#1}{\urlprefix }}%
\providecommand \urlprefix  [0]{URL }%
\providecommand \Eprint [0]{\href }%
\providecommand \doibase [0]{http://dx.doi.org/}%
\providecommand \selectlanguage [0]{\@gobble}%
\providecommand \bibinfo  [0]{\@secondoftwo}%
\providecommand \bibfield  [0]{\@secondoftwo}%
\providecommand \translation [1]{[#1]}%
\providecommand \BibitemOpen [0]{}%
\providecommand \bibitemStop [0]{}%
\providecommand \bibitemNoStop [0]{.\EOS\space}%
\providecommand \EOS [0]{\spacefactor3000\relax}%
\providecommand \BibitemShut  [1]{\csname bibitem#1\endcsname}%
\let\auto@bib@innerbib\@empty
\bibitem [{\citenamefont {Watson}\ \emph {et~al.}(2018)\citenamefont {Watson},
  \citenamefont {Philips}, \citenamefont {Kawakami}, \citenamefont {Ward},
  \citenamefont {Scarlino}, \citenamefont {Veldhorst}, \citenamefont {Savage},
  \citenamefont {Lagally}, \citenamefont {Friesen}, \citenamefont
  {Coppersmith}, \citenamefont {Eriksson},\ and\ \citenamefont
  {Vandersypen}}]{watson}%
  \BibitemOpen
  \bibfield  {author} {\bibinfo {author} {\bibfnamefont {T.~F.}\ \bibnamefont
  {Watson}}, \bibinfo {author} {\bibfnamefont {S.~G.}\ \bibnamefont {Philips}},
  \bibinfo {author} {\bibfnamefont {E.}~\bibnamefont {Kawakami}}, \bibinfo
  {author} {\bibfnamefont {D.~R.}\ \bibnamefont {Ward}}, \bibinfo {author}
  {\bibfnamefont {P.}~\bibnamefont {Scarlino}}, \bibinfo {author}
  {\bibfnamefont {M.}~\bibnamefont {Veldhorst}}, \bibinfo {author}
  {\bibfnamefont {D.~E.}\ \bibnamefont {Savage}}, \bibinfo {author}
  {\bibfnamefont {M.~G.}\ \bibnamefont {Lagally}}, \bibinfo {author}
  {\bibfnamefont {M.}~\bibnamefont {Friesen}}, \bibinfo {author} {\bibfnamefont
  {S.~N.}\ \bibnamefont {Coppersmith}}, \bibinfo {author} {\bibfnamefont
  {M.~A.}\ \bibnamefont {Eriksson}}, \ and\ \bibinfo {author} {\bibfnamefont
  {L.~M.}\ \bibnamefont {Vandersypen}},\ }\href {\doibase 10.1038/nature25766}
  {\bibfield  {journal} {\bibinfo  {journal} {Nature}\ }\textbf {\bibinfo
  {volume} {555}},\ \bibinfo {pages} {633} (\bibinfo {year}
  {2018})}\BibitemShut {NoStop}%
\bibitem [{\citenamefont {Kawakami}\ \emph {et~al.}(2014)\citenamefont
  {Kawakami}, \citenamefont {Scarlino}, \citenamefont {Ward}, \citenamefont
  {Braakman}, \citenamefont {Savage}, \citenamefont {Lagally}, \citenamefont
  {Friesen}, \citenamefont {Coppersmith}, \citenamefont {Eriksson},\ and\
  \citenamefont {Vandersypen}}]{Kawakami2014}%
  \BibitemOpen
  \bibfield  {author} {\bibinfo {author} {\bibfnamefont {E.}~\bibnamefont
  {Kawakami}}, \bibinfo {author} {\bibfnamefont {P.}~\bibnamefont {Scarlino}},
  \bibinfo {author} {\bibfnamefont {D.~R.}\ \bibnamefont {Ward}}, \bibinfo
  {author} {\bibfnamefont {F.~R.}\ \bibnamefont {Braakman}}, \bibinfo {author}
  {\bibfnamefont {D.~E.}\ \bibnamefont {Savage}}, \bibinfo {author}
  {\bibfnamefont {M.~G.}\ \bibnamefont {Lagally}}, \bibinfo {author}
  {\bibfnamefont {M.}~\bibnamefont {Friesen}}, \bibinfo {author} {\bibfnamefont
  {S.~N.}\ \bibnamefont {Coppersmith}}, \bibinfo {author} {\bibfnamefont
  {M.~A.}\ \bibnamefont {Eriksson}}, \ and\ \bibinfo {author} {\bibfnamefont
  {L.~M.}\ \bibnamefont {Vandersypen}},\ }\href {\doibase
  10.1038/nnano.2014.153} {\bibfield  {journal} {\bibinfo  {journal} {Nature
  Nanotechnology}\ }\textbf {\bibinfo {volume} {9}},\ \bibinfo {pages} {666}
  (\bibinfo {year} {2014})}\BibitemShut {NoStop}%
\bibitem [{\citenamefont {Takeda}\ \emph {et~al.}(2016)\citenamefont {Takeda},
  \citenamefont {Kamioka}, \citenamefont {Otsuka}, \citenamefont {Yoneda},
  \citenamefont {Nakajima}, \citenamefont {Delbecq}, \citenamefont {Amaha},
  \citenamefont {Allison}, \citenamefont {Kodera}, \citenamefont {Oda} \emph
  {et~al.}}]{takeda2016fault}%
  \BibitemOpen
  \bibfield  {author} {\bibinfo {author} {\bibfnamefont {K.}~\bibnamefont
  {Takeda}}, \bibinfo {author} {\bibfnamefont {J.}~\bibnamefont {Kamioka}},
  \bibinfo {author} {\bibfnamefont {T.}~\bibnamefont {Otsuka}}, \bibinfo
  {author} {\bibfnamefont {J.}~\bibnamefont {Yoneda}}, \bibinfo {author}
  {\bibfnamefont {T.}~\bibnamefont {Nakajima}}, \bibinfo {author}
  {\bibfnamefont {M.~R.}\ \bibnamefont {Delbecq}}, \bibinfo {author}
  {\bibfnamefont {S.}~\bibnamefont {Amaha}}, \bibinfo {author} {\bibfnamefont
  {G.}~\bibnamefont {Allison}}, \bibinfo {author} {\bibfnamefont
  {T.}~\bibnamefont {Kodera}}, \bibinfo {author} {\bibfnamefont
  {S.}~\bibnamefont {Oda}},  \emph {et~al.},\ }\href@noop {} {\bibfield
  {journal} {\bibinfo  {journal} {Science Advances}\ }\textbf {\bibinfo
  {volume} {2}},\ \bibinfo {pages} {e1600694} (\bibinfo {year}
  {2016})}\BibitemShut {NoStop}%
\bibitem [{\citenamefont {Pioro-Ladri{\`{e}}re}\ \emph
  {et~al.}(2008)\citenamefont {Pioro-Ladri{\`{e}}re}, \citenamefont {Obata},
  \citenamefont {Tokura}, \citenamefont {Shin}, \citenamefont {Kubo},
  \citenamefont {Yoshida}, \citenamefont {Taniyama},\ and\ \citenamefont
  {Tarucha}}]{Pioro-Ladriere2008}%
  \BibitemOpen
  \bibfield  {author} {\bibinfo {author} {\bibfnamefont {M.}~\bibnamefont
  {Pioro-Ladri{\`{e}}re}}, \bibinfo {author} {\bibfnamefont {T.}~\bibnamefont
  {Obata}}, \bibinfo {author} {\bibfnamefont {Y.}~\bibnamefont {Tokura}},
  \bibinfo {author} {\bibfnamefont {Y.~S.}\ \bibnamefont {Shin}}, \bibinfo
  {author} {\bibfnamefont {T.}~\bibnamefont {Kubo}}, \bibinfo {author}
  {\bibfnamefont {K.}~\bibnamefont {Yoshida}}, \bibinfo {author} {\bibfnamefont
  {T.}~\bibnamefont {Taniyama}}, \ and\ \bibinfo {author} {\bibfnamefont
  {S.}~\bibnamefont {Tarucha}},\ }\href {\doibase 10.1038/NPHYS1053} {\bibfield
   {journal} {\bibinfo  {journal} {Nature Physics}\ }\textbf {\bibinfo {volume}
  {4}},\ \bibinfo {pages} {776} (\bibinfo {year} {2008})}\BibitemShut {NoStop}%
\bibitem [{\citenamefont {Philips}\ \emph {et~al.}(2022)\citenamefont
  {Philips}, \citenamefont {Madzik}, \citenamefont {Amitonov}, \citenamefont
  {de~Snoo}, \citenamefont {Russ}, \citenamefont {Kalhor}, \citenamefont
  {Volk}, \citenamefont {Lawrie}, \citenamefont {Brousse}, \citenamefont
  {Tryputen}, \citenamefont {Wuetz}, \citenamefont {Sammak}, \citenamefont
  {Veldhorst}, \citenamefont {Scappucci},\ and\ \citenamefont
  {Vandersypen}}]{philips2022universal}%
  \BibitemOpen
  \bibfield  {author} {\bibinfo {author} {\bibfnamefont {S.~G.}\ \bibnamefont
  {Philips}}, \bibinfo {author} {\bibfnamefont {M.~T.}\ \bibnamefont {Madzik}},
  \bibinfo {author} {\bibfnamefont {S.~V.}\ \bibnamefont {Amitonov}}, \bibinfo
  {author} {\bibfnamefont {S.~L.}\ \bibnamefont {de~Snoo}}, \bibinfo {author}
  {\bibfnamefont {M.}~\bibnamefont {Russ}}, \bibinfo {author} {\bibfnamefont
  {N.}~\bibnamefont {Kalhor}}, \bibinfo {author} {\bibfnamefont
  {C.}~\bibnamefont {Volk}}, \bibinfo {author} {\bibfnamefont {W.~I.}\
  \bibnamefont {Lawrie}}, \bibinfo {author} {\bibfnamefont {D.}~\bibnamefont
  {Brousse}}, \bibinfo {author} {\bibfnamefont {L.}~\bibnamefont {Tryputen}},
  \bibinfo {author} {\bibfnamefont {B.~P.}\ \bibnamefont {Wuetz}}, \bibinfo
  {author} {\bibfnamefont {A.}~\bibnamefont {Sammak}}, \bibinfo {author}
  {\bibfnamefont {M.}~\bibnamefont {Veldhorst}}, \bibinfo {author}
  {\bibfnamefont {G.}~\bibnamefont {Scappucci}}, \ and\ \bibinfo {author}
  {\bibfnamefont {L.~M.}\ \bibnamefont {Vandersypen}},\ }\href@noop {}
  {\bibfield  {journal} {\bibinfo  {journal} {Nature}\ }\textbf {\bibinfo
  {volume} {4}},\ \bibinfo {pages} {919} (\bibinfo {year} {2022})}\BibitemShut
  {NoStop}%
\bibitem [{\citenamefont {Gonzalez-Zalba}\ \emph {et~al.}(2021)\citenamefont
  {Gonzalez-Zalba}, \citenamefont {de~Franceschi}, \citenamefont {Charbon},
  \citenamefont {Meunier}, \citenamefont {Vinet},\ and\ \citenamefont
  {Dzurak}}]{gonzalez2021scaling}%
  \BibitemOpen
  \bibfield  {author} {\bibinfo {author} {\bibfnamefont {M.}~\bibnamefont
  {Gonzalez-Zalba}}, \bibinfo {author} {\bibfnamefont {S.}~\bibnamefont
  {de~Franceschi}}, \bibinfo {author} {\bibfnamefont {E.}~\bibnamefont
  {Charbon}}, \bibinfo {author} {\bibfnamefont {T.}~\bibnamefont {Meunier}},
  \bibinfo {author} {\bibfnamefont {M.}~\bibnamefont {Vinet}}, \ and\ \bibinfo
  {author} {\bibfnamefont {A.}~\bibnamefont {Dzurak}},\ }\href@noop {}
  {\bibfield  {journal} {\bibinfo  {journal} {Nature Electronics}\ }\textbf
  {\bibinfo {volume} {4}},\ \bibinfo {pages} {872} (\bibinfo {year}
  {2021})}\BibitemShut {NoStop}%
\bibitem [{\citenamefont {Tadokoro}\ \emph {et~al.}(2021)\citenamefont
  {Tadokoro}, \citenamefont {Nakajima}, \citenamefont {Kobayashi},
  \citenamefont {Takeda}, \citenamefont {Noiri}, \citenamefont {Tomari},
  \citenamefont {Yoneda}, \citenamefont {Tarucha},\ and\ \citenamefont
  {Kodera}}]{tadokoro2021designs}%
  \BibitemOpen
  \bibfield  {author} {\bibinfo {author} {\bibfnamefont {M.}~\bibnamefont
  {Tadokoro}}, \bibinfo {author} {\bibfnamefont {T.}~\bibnamefont {Nakajima}},
  \bibinfo {author} {\bibfnamefont {T.}~\bibnamefont {Kobayashi}}, \bibinfo
  {author} {\bibfnamefont {K.}~\bibnamefont {Takeda}}, \bibinfo {author}
  {\bibfnamefont {A.}~\bibnamefont {Noiri}}, \bibinfo {author} {\bibfnamefont
  {K.}~\bibnamefont {Tomari}}, \bibinfo {author} {\bibfnamefont
  {J.}~\bibnamefont {Yoneda}}, \bibinfo {author} {\bibfnamefont
  {S.}~\bibnamefont {Tarucha}}, \ and\ \bibinfo {author} {\bibfnamefont
  {T.}~\bibnamefont {Kodera}},\ }\href@noop {} {\bibfield  {journal} {\bibinfo
  {journal} {Scientific Reports}\ }\textbf {\bibinfo {volume} {11}},\ \bibinfo
  {pages} {19406} (\bibinfo {year} {2021})}\BibitemShut {NoStop}%
\bibitem [{\citenamefont {Boter}\ \emph {et~al.}(2022)\citenamefont {Boter},
  \citenamefont {Dehollain}, \citenamefont {van Dijk}, \citenamefont {Xu},
  \citenamefont {Hensgens}, \citenamefont {Versluis}, \citenamefont {Naus},
  \citenamefont {Clarke}, \citenamefont {Veldhorst}, \citenamefont
  {Sebastiano},\ and\ \citenamefont {Vandersypen}}]{boter2021spider}%
  \BibitemOpen
  \bibfield  {author} {\bibinfo {author} {\bibfnamefont {J.~M.}\ \bibnamefont
  {Boter}}, \bibinfo {author} {\bibfnamefont {J.~P.}\ \bibnamefont
  {Dehollain}}, \bibinfo {author} {\bibfnamefont {J.~P.}\ \bibnamefont {van
  Dijk}}, \bibinfo {author} {\bibfnamefont {Y.}~\bibnamefont {Xu}}, \bibinfo
  {author} {\bibfnamefont {T.}~\bibnamefont {Hensgens}}, \bibinfo {author}
  {\bibfnamefont {R.}~\bibnamefont {Versluis}}, \bibinfo {author}
  {\bibfnamefont {H.~W.}\ \bibnamefont {Naus}}, \bibinfo {author}
  {\bibfnamefont {J.~S.}\ \bibnamefont {Clarke}}, \bibinfo {author}
  {\bibfnamefont {M.}~\bibnamefont {Veldhorst}}, \bibinfo {author}
  {\bibfnamefont {F.}~\bibnamefont {Sebastiano}}, \ and\ \bibinfo {author}
  {\bibfnamefont {L.~M.}\ \bibnamefont {Vandersypen}},\ }\href {\doibase
  10.1103/PhysRevApplied.18.024053} {\bibfield  {journal} {\bibinfo  {journal}
  {Phys. Rev. Appl.}\ }\textbf {\bibinfo {volume} {18}},\ \bibinfo {pages}
  {024053} (\bibinfo {year} {2022})}\BibitemShut {NoStop}%
\bibitem [{\citenamefont {Nakamura}\ \emph {et~al.}(2022)\citenamefont
  {Nakamura}, \citenamefont {Kiyama},\ and\ \citenamefont
  {Oiwa}}]{nakamura2022micromagnet}%
  \BibitemOpen
  \bibfield  {author} {\bibinfo {author} {\bibfnamefont {S.}~\bibnamefont
  {Nakamura}}, \bibinfo {author} {\bibfnamefont {H.}~\bibnamefont {Kiyama}}, \
  and\ \bibinfo {author} {\bibfnamefont {A.}~\bibnamefont {Oiwa}},\ }\href@noop
  {} {\bibfield  {journal} {\bibinfo  {journal} {Journal of Applied Physics}\
  }\textbf {\bibinfo {volume} {132}},\ \bibinfo {pages} {224301} (\bibinfo
  {year} {2022})}\BibitemShut {NoStop}%
\bibitem [{\citenamefont {Yoneda}\ \emph {et~al.}(2015)\citenamefont {Yoneda},
  \citenamefont {Otsuka}, \citenamefont {Takakura}, \citenamefont
  {Pioro-Ladri{\`e}re}, \citenamefont {Brunner}, \citenamefont {Lu},
  \citenamefont {Nakajima}, \citenamefont {Obata}, \citenamefont {Noiri},
  \citenamefont {Palmstr{\o}m} \emph {et~al.}}]{yoneda2015robust}%
  \BibitemOpen
  \bibfield  {author} {\bibinfo {author} {\bibfnamefont {J.}~\bibnamefont
  {Yoneda}}, \bibinfo {author} {\bibfnamefont {T.}~\bibnamefont {Otsuka}},
  \bibinfo {author} {\bibfnamefont {T.}~\bibnamefont {Takakura}}, \bibinfo
  {author} {\bibfnamefont {M.}~\bibnamefont {Pioro-Ladri{\`e}re}}, \bibinfo
  {author} {\bibfnamefont {R.}~\bibnamefont {Brunner}}, \bibinfo {author}
  {\bibfnamefont {H.}~\bibnamefont {Lu}}, \bibinfo {author} {\bibfnamefont
  {T.}~\bibnamefont {Nakajima}}, \bibinfo {author} {\bibfnamefont
  {T.}~\bibnamefont {Obata}}, \bibinfo {author} {\bibfnamefont
  {A.}~\bibnamefont {Noiri}}, \bibinfo {author} {\bibfnamefont {C.~J.}\
  \bibnamefont {Palmstr{\o}m}},  \emph {et~al.},\ }\href@noop {} {\bibfield
  {journal} {\bibinfo  {journal} {Applied Physics Express}\ }\textbf {\bibinfo
  {volume} {8}},\ \bibinfo {pages} {084401} (\bibinfo {year}
  {2015})}\BibitemShut {NoStop}%
\bibitem [{\citenamefont {Froning}\ \emph {et~al.}(2018)\citenamefont
  {Froning}, \citenamefont {Rehmann}, \citenamefont {Ridderbos}, \citenamefont
  {Brauns}, \citenamefont {Zwanenburg}, \citenamefont {Li}, \citenamefont
  {Bakkers}, \citenamefont {Zumbühl},\ and\ \citenamefont
  {Braakman}}]{froning2018single}%
  \BibitemOpen
  \bibfield  {author} {\bibinfo {author} {\bibfnamefont {F.~N.~M.}\
  \bibnamefont {Froning}}, \bibinfo {author} {\bibfnamefont {M.~K.}\
  \bibnamefont {Rehmann}}, \bibinfo {author} {\bibfnamefont {J.}~\bibnamefont
  {Ridderbos}}, \bibinfo {author} {\bibfnamefont {M.}~\bibnamefont {Brauns}},
  \bibinfo {author} {\bibfnamefont {F.~A.}\ \bibnamefont {Zwanenburg}},
  \bibinfo {author} {\bibfnamefont {A.}~\bibnamefont {Li}}, \bibinfo {author}
  {\bibfnamefont {E.~P. A.~M.}\ \bibnamefont {Bakkers}}, \bibinfo {author}
  {\bibfnamefont {D.~M.}\ \bibnamefont {Zumbühl}}, \ and\ \bibinfo {author}
  {\bibfnamefont {F.~R.}\ \bibnamefont {Braakman}},\ }\href {\doibase
  10.1063/1.5042501} {\bibfield  {journal} {\bibinfo  {journal} {Applied
  Physics Letters}\ }\textbf {\bibinfo {volume} {113}},\ \bibinfo {pages}
  {073102} (\bibinfo {year} {2018})},\ \Eprint
  {http://arxiv.org/abs/https://doi.org/10.1063/1.5042501}
  {https://doi.org/10.1063/1.5042501} \BibitemShut {NoStop}%
\bibitem [{\citenamefont {Kuhlmann}\ \emph {et~al.}(2018)\citenamefont
  {Kuhlmann}, \citenamefont {Deshpande}, \citenamefont {Camenzind},
  \citenamefont {Zumb{\"u}hl},\ and\ \citenamefont
  {Fuhrer}}]{kuhlmann2018ambipolar}%
  \BibitemOpen
  \bibfield  {author} {\bibinfo {author} {\bibfnamefont {A.~V.}\ \bibnamefont
  {Kuhlmann}}, \bibinfo {author} {\bibfnamefont {V.}~\bibnamefont {Deshpande}},
  \bibinfo {author} {\bibfnamefont {L.~C.}\ \bibnamefont {Camenzind}}, \bibinfo
  {author} {\bibfnamefont {D.~M.}\ \bibnamefont {Zumb{\"u}hl}}, \ and\ \bibinfo
  {author} {\bibfnamefont {A.}~\bibnamefont {Fuhrer}},\ }\href@noop {}
  {\bibfield  {journal} {\bibinfo  {journal} {Applied Physics Letters}\
  }\textbf {\bibinfo {volume} {113}},\ \bibinfo {pages} {122107} (\bibinfo
  {year} {2018})}\BibitemShut {NoStop}%
\bibitem [{\citenamefont {Zwerver}\ \emph {et~al.}(2022)\citenamefont
  {Zwerver}, \citenamefont {Kr{\"a}henmann}, \citenamefont {Watson},
  \citenamefont {Lampert}, \citenamefont {George}, \citenamefont
  {Pillarisetty}, \citenamefont {Bojarski}, \citenamefont {Amin}, \citenamefont
  {Amitonov}, \citenamefont {Boter} \emph {et~al.}}]{zwerver2022qubits}%
  \BibitemOpen
  \bibfield  {author} {\bibinfo {author} {\bibfnamefont {A.}~\bibnamefont
  {Zwerver}}, \bibinfo {author} {\bibfnamefont {T.}~\bibnamefont
  {Kr{\"a}henmann}}, \bibinfo {author} {\bibfnamefont {T.}~\bibnamefont
  {Watson}}, \bibinfo {author} {\bibfnamefont {L.}~\bibnamefont {Lampert}},
  \bibinfo {author} {\bibfnamefont {H.~C.}\ \bibnamefont {George}}, \bibinfo
  {author} {\bibfnamefont {R.}~\bibnamefont {Pillarisetty}}, \bibinfo {author}
  {\bibfnamefont {S.}~\bibnamefont {Bojarski}}, \bibinfo {author}
  {\bibfnamefont {P.}~\bibnamefont {Amin}}, \bibinfo {author} {\bibfnamefont
  {S.}~\bibnamefont {Amitonov}}, \bibinfo {author} {\bibfnamefont
  {J.}~\bibnamefont {Boter}},  \emph {et~al.},\ }\href@noop {} {\bibfield
  {journal} {\bibinfo  {journal} {Nature Electronics}\ }\textbf {\bibinfo
  {volume} {5}},\ \bibinfo {pages} {184} (\bibinfo {year} {2022})}\BibitemShut
  {NoStop}%
\bibitem [{\citenamefont {Allenspach}(2000)}]{allenspach2000spin}%
  \BibitemOpen
  \bibfield  {author} {\bibinfo {author} {\bibfnamefont {R.}~\bibnamefont
  {Allenspach}},\ }\href@noop {} {\bibfield  {journal} {\bibinfo  {journal}
  {IBM Journal of Research and Development}\ }\textbf {\bibinfo {volume}
  {44}},\ \bibinfo {pages} {553} (\bibinfo {year} {2000})}\BibitemShut
  {NoStop}%
\bibitem [{\citenamefont {Vansteenkiste}\ \emph {et~al.}(2014)\citenamefont
  {Vansteenkiste}, \citenamefont {Leliaert}, \citenamefont {Dvornik},
  \citenamefont {Helsen}, \citenamefont {Garcia-Sanchez},\ and\ \citenamefont
  {Van~Waeyenberge}}]{vansteenkiste2014design}%
  \BibitemOpen
  \bibfield  {author} {\bibinfo {author} {\bibfnamefont {A.}~\bibnamefont
  {Vansteenkiste}}, \bibinfo {author} {\bibfnamefont {J.}~\bibnamefont
  {Leliaert}}, \bibinfo {author} {\bibfnamefont {M.}~\bibnamefont {Dvornik}},
  \bibinfo {author} {\bibfnamefont {M.}~\bibnamefont {Helsen}}, \bibinfo
  {author} {\bibfnamefont {F.}~\bibnamefont {Garcia-Sanchez}}, \ and\ \bibinfo
  {author} {\bibfnamefont {B.}~\bibnamefont {Van~Waeyenberge}},\ }\href@noop {}
  {\bibfield  {journal} {\bibinfo  {journal} {AIP Advances}\ }\textbf {\bibinfo
  {volume} {4}},\ \bibinfo {pages} {107133} (\bibinfo {year}
  {2014})}\BibitemShut {NoStop}%
\bibitem [{\citenamefont {Struck}\ \emph {et~al.}(2020)\citenamefont {Struck},
  \citenamefont {Hollmann}, \citenamefont {Schauer}, \citenamefont {Fedorets},
  \citenamefont {Schmidbauer}, \citenamefont {Sawano}, \citenamefont {Riemann},
  \citenamefont {Abrosimov}, \citenamefont {Cywi{\'n}ski}, \citenamefont
  {Bougeard} \emph {et~al.}}]{struck2020low}%
  \BibitemOpen
  \bibfield  {author} {\bibinfo {author} {\bibfnamefont {T.}~\bibnamefont
  {Struck}}, \bibinfo {author} {\bibfnamefont {A.}~\bibnamefont {Hollmann}},
  \bibinfo {author} {\bibfnamefont {F.}~\bibnamefont {Schauer}}, \bibinfo
  {author} {\bibfnamefont {O.}~\bibnamefont {Fedorets}}, \bibinfo {author}
  {\bibfnamefont {A.}~\bibnamefont {Schmidbauer}}, \bibinfo {author}
  {\bibfnamefont {K.}~\bibnamefont {Sawano}}, \bibinfo {author} {\bibfnamefont
  {H.}~\bibnamefont {Riemann}}, \bibinfo {author} {\bibfnamefont {N.~V.}\
  \bibnamefont {Abrosimov}}, \bibinfo {author} {\bibfnamefont
  {{\L}.}~\bibnamefont {Cywi{\'n}ski}}, \bibinfo {author} {\bibfnamefont
  {D.}~\bibnamefont {Bougeard}},  \emph {et~al.},\ }\href@noop {} {\bibfield
  {journal} {\bibinfo  {journal} {npj Quantum Information}\ }\textbf {\bibinfo
  {volume} {6}},\ \bibinfo {pages} {1} (\bibinfo {year} {2020})}\BibitemShut
  {NoStop}%
\bibitem [{\citenamefont {Yoneda}\ \emph {et~al.}(2018)\citenamefont {Yoneda},
  \citenamefont {Takeda}, \citenamefont {Otsuka}, \citenamefont {Nakajima},
  \citenamefont {Delbecq}, \citenamefont {Allison}, \citenamefont {Honda},
  \citenamefont {Kodera}, \citenamefont {Oda}, \citenamefont {Hoshi} \emph
  {et~al.}}]{yoneda2018quantum}%
  \BibitemOpen
  \bibfield  {author} {\bibinfo {author} {\bibfnamefont {J.}~\bibnamefont
  {Yoneda}}, \bibinfo {author} {\bibfnamefont {K.}~\bibnamefont {Takeda}},
  \bibinfo {author} {\bibfnamefont {T.}~\bibnamefont {Otsuka}}, \bibinfo
  {author} {\bibfnamefont {T.}~\bibnamefont {Nakajima}}, \bibinfo {author}
  {\bibfnamefont {M.~R.}\ \bibnamefont {Delbecq}}, \bibinfo {author}
  {\bibfnamefont {G.}~\bibnamefont {Allison}}, \bibinfo {author} {\bibfnamefont
  {T.}~\bibnamefont {Honda}}, \bibinfo {author} {\bibfnamefont
  {T.}~\bibnamefont {Kodera}}, \bibinfo {author} {\bibfnamefont
  {S.}~\bibnamefont {Oda}}, \bibinfo {author} {\bibfnamefont {Y.}~\bibnamefont
  {Hoshi}},  \emph {et~al.},\ }\href {\doibase 10.1038/s41565-017-0014-x}
  {\bibfield  {journal} {\bibinfo  {journal} {Nature Nanotechnology}\ }\textbf
  {\bibinfo {volume} {13}},\ \bibinfo {pages} {102} (\bibinfo {year}
  {2018})}\BibitemShut {NoStop}%
\bibitem [{\citenamefont {Neumann}\ and\ \citenamefont
  {Schreiber}(2015)}]{neumann2015simulation}%
  \BibitemOpen
  \bibfield  {author} {\bibinfo {author} {\bibfnamefont {R.}~\bibnamefont
  {Neumann}}\ and\ \bibinfo {author} {\bibfnamefont {L.}~\bibnamefont
  {Schreiber}},\ }\href@noop {} {\bibfield  {journal} {\bibinfo  {journal}
  {Journal of Applied Physics}\ }\textbf {\bibinfo {volume} {117}},\ \bibinfo
  {pages} {193903} (\bibinfo {year} {2015})}\BibitemShut {NoStop}%
\bibitem [{\citenamefont {Chesi}\ \emph {et~al.}(2014)\citenamefont {Chesi},
  \citenamefont {Wang}, \citenamefont {Yoneda}, \citenamefont {Otsuka},
  \citenamefont {Tarucha},\ and\ \citenamefont {Loss}}]{chesi2014single}%
  \BibitemOpen
  \bibfield  {author} {\bibinfo {author} {\bibfnamefont {S.}~\bibnamefont
  {Chesi}}, \bibinfo {author} {\bibfnamefont {Y.-D.}\ \bibnamefont {Wang}},
  \bibinfo {author} {\bibfnamefont {J.}~\bibnamefont {Yoneda}}, \bibinfo
  {author} {\bibfnamefont {T.}~\bibnamefont {Otsuka}}, \bibinfo {author}
  {\bibfnamefont {S.}~\bibnamefont {Tarucha}}, \ and\ \bibinfo {author}
  {\bibfnamefont {D.}~\bibnamefont {Loss}},\ }\href@noop {} {\bibfield
  {journal} {\bibinfo  {journal} {Physical Review B}\ }\textbf {\bibinfo
  {volume} {90}},\ \bibinfo {pages} {235311} (\bibinfo {year}
  {2014})}\BibitemShut {NoStop}%
\bibitem [{\citenamefont {Teske}\ \emph {et~al.}(2022)\citenamefont {Teske},
  \citenamefont {Butt}, \citenamefont {Cerfontaine}, \citenamefont {Burkard},\
  and\ \citenamefont {Bluhm}}]{tesker2022flopping}%
  \BibitemOpen
  \bibfield  {author} {\bibinfo {author} {\bibfnamefont {J.~D.}\ \bibnamefont
  {Teske}}, \bibinfo {author} {\bibfnamefont {F.}~\bibnamefont {Butt}},
  \bibinfo {author} {\bibfnamefont {P.}~\bibnamefont {Cerfontaine}}, \bibinfo
  {author} {\bibfnamefont {G.}~\bibnamefont {Burkard}}, \ and\ \bibinfo
  {author} {\bibfnamefont {H.}~\bibnamefont {Bluhm}},\ }\href@noop {}
  {\bibfield  {journal} {\bibinfo  {journal} {arXiv preprint arXiv:2208.10548}\
  } (\bibinfo {year} {2022})}\BibitemShut {NoStop}%
\bibitem [{\citenamefont {Glavin}\ and\ \citenamefont
  {Kim}(2003)}]{glavin2003spin}%
  \BibitemOpen
  \bibfield  {author} {\bibinfo {author} {\bibfnamefont {B.}~\bibnamefont
  {Glavin}}\ and\ \bibinfo {author} {\bibfnamefont {K.}~\bibnamefont {Kim}},\
  }\href@noop {} {\bibfield  {journal} {\bibinfo  {journal} {Physical Review
  B}\ }\textbf {\bibinfo {volume} {68}},\ \bibinfo {pages} {045308} (\bibinfo
  {year} {2003})}\BibitemShut {NoStop}%
\bibitem [{\citenamefont {Huang}\ and\ \citenamefont
  {Hu}(2014)}]{huang2014spin}%
  \BibitemOpen
  \bibfield  {author} {\bibinfo {author} {\bibfnamefont {P.}~\bibnamefont
  {Huang}}\ and\ \bibinfo {author} {\bibfnamefont {X.}~\bibnamefont {Hu}},\
  }\href@noop {} {\bibfield  {journal} {\bibinfo  {journal} {Physical Review
  B}\ }\textbf {\bibinfo {volume} {90}},\ \bibinfo {pages} {235315} (\bibinfo
  {year} {2014})}\BibitemShut {NoStop}%
\bibitem [{\citenamefont {Langrock}\ \emph {et~al.}(2022)\citenamefont
  {Langrock}, \citenamefont {Krzywda}, \citenamefont {Focke}, \citenamefont
  {Seidler}, \citenamefont {Schreiber},\ and\ \citenamefont
  {Cywi{\'n}ski}}]{langrock2022blueprint}%
  \BibitemOpen
  \bibfield  {author} {\bibinfo {author} {\bibfnamefont {V.}~\bibnamefont
  {Langrock}}, \bibinfo {author} {\bibfnamefont {J.~A.}\ \bibnamefont
  {Krzywda}}, \bibinfo {author} {\bibfnamefont {N.}~\bibnamefont {Focke}},
  \bibinfo {author} {\bibfnamefont {I.}~\bibnamefont {Seidler}}, \bibinfo
  {author} {\bibfnamefont {L.~R.}\ \bibnamefont {Schreiber}}, \ and\ \bibinfo
  {author} {\bibfnamefont {{\L}.}~\bibnamefont {Cywi{\'n}ski}},\ }\href@noop {}
  {\bibfield  {journal} {\bibinfo  {journal} {arXiv preprint arXiv:2202.11793}\
  } (\bibinfo {year} {2022})}\BibitemShut {NoStop}%
\bibitem [{\citenamefont {Nakajima}\ \emph {et~al.}(2020)\citenamefont
  {Nakajima}, \citenamefont {Noiri}, \citenamefont {Kawasaki}, \citenamefont
  {Yoneda}, \citenamefont {Stano}, \citenamefont {Amaha}, \citenamefont
  {Otsuka}, \citenamefont {Takeda}, \citenamefont {Delbecq}, \citenamefont
  {Allison} \emph {et~al.}}]{nakajima2020coherence}%
  \BibitemOpen
  \bibfield  {author} {\bibinfo {author} {\bibfnamefont {T.}~\bibnamefont
  {Nakajima}}, \bibinfo {author} {\bibfnamefont {A.}~\bibnamefont {Noiri}},
  \bibinfo {author} {\bibfnamefont {K.}~\bibnamefont {Kawasaki}}, \bibinfo
  {author} {\bibfnamefont {J.}~\bibnamefont {Yoneda}}, \bibinfo {author}
  {\bibfnamefont {P.}~\bibnamefont {Stano}}, \bibinfo {author} {\bibfnamefont
  {S.}~\bibnamefont {Amaha}}, \bibinfo {author} {\bibfnamefont
  {T.}~\bibnamefont {Otsuka}}, \bibinfo {author} {\bibfnamefont
  {K.}~\bibnamefont {Takeda}}, \bibinfo {author} {\bibfnamefont {M.~R.}\
  \bibnamefont {Delbecq}}, \bibinfo {author} {\bibfnamefont {G.}~\bibnamefont
  {Allison}},  \emph {et~al.},\ }\href@noop {} {\bibfield  {journal} {\bibinfo
  {journal} {Physical Review X}\ }\textbf {\bibinfo {volume} {10}},\ \bibinfo
  {pages} {011060} (\bibinfo {year} {2020})}\BibitemShut {NoStop}%
\bibitem [{\citenamefont {Dumoulin~Stuyck}\ \emph {et~al.}(2021)\citenamefont
  {Dumoulin~Stuyck}, \citenamefont {Mohiyaddin}, \citenamefont {Li},
  \citenamefont {Heyns}, \citenamefont {Govoreanu},\ and\ \citenamefont
  {Radu}}]{dumoulin2021low}%
  \BibitemOpen
  \bibfield  {author} {\bibinfo {author} {\bibfnamefont {N.~I.}\ \bibnamefont
  {Dumoulin~Stuyck}}, \bibinfo {author} {\bibfnamefont {F.~A.}\ \bibnamefont
  {Mohiyaddin}}, \bibinfo {author} {\bibfnamefont {R.}~\bibnamefont {Li}},
  \bibinfo {author} {\bibfnamefont {M.}~\bibnamefont {Heyns}}, \bibinfo
  {author} {\bibfnamefont {B.}~\bibnamefont {Govoreanu}}, \ and\ \bibinfo
  {author} {\bibfnamefont {I.~P.}\ \bibnamefont {Radu}},\ }\href {\doibase
  10.1063/5.0059939} {\bibfield  {journal} {\bibinfo  {journal} {Applied
  Physics Letters}\ }\textbf {\bibinfo {volume} {119}},\ \bibinfo {pages}
  {094001} (\bibinfo {year} {2021})},\ \Eprint
  {http://arxiv.org/abs/https://doi.org/10.1063/5.0059939}
  {https://doi.org/10.1063/5.0059939} \BibitemShut {NoStop}%
\bibitem [{\citenamefont {Raith}(2015)}]{ebl}%
  \BibitemOpen
  \bibfield  {author} {\bibinfo {author} {\bibnamefont {Raith}},\ }\href@noop
  {} {\enquote {\bibinfo {title} {Datasheet ebpg5200 plus},}\ }\bibinfo
  {howpublished} {\url{https://raith.com/product/ebpg-plus/}} (\bibinfo {year}
  {2015}),\ \bibinfo {note} {accessed: 2022-12-01}\BibitemShut {NoStop}%
\end{thebibliography}%

\end{document}